\documentclass[aps,prb,twocolumn,showpacs,floatfix]{revtex4}
\usepackage{graphicx}
\usepackage{dcolumn}
\usepackage{amsmath,mathrsfs,amssymb,amsthm}
\usepackage{hhline}
\usepackage{wasysym,enumerate,color}
\newcommand{\be}{\begin{equation}}
\newcommand{\ee}{\end{equation}}
\newcommand{\bea}{\begin{eqnarray}}
\newcommand{\eea}{\end{eqnarray}}

\newcommand{\ds}{\displaystyle}
\newcommand{\1}{{1\hspace*{-0.5ex} \textrm{l} \hspace*{0.5ex}}}
\newcommand{\tr}{{\rm Tr\;}}

\begin{document}
\title{Density matrix numerical renormalization group for non-Abelian
symmetries}
\author{A.\ I.\ T\'oth,$^{a,b}$
C.\ P.\ Moca,$^{a,c}$ \"O.\ Legeza$^{a,d}$ 
and G.\ Zar\'and$^{a}$}
\affiliation{
$^a$ Theoretical Physics Department, Institute of Physics, 
Budapest University of Technology and Economics, H-1521 Budapest, Hungary\\
$^b$ Institute for Theoretische Festk\"orper Physik, Universit\"at Karlsruhe,
  D-76128 Karlsruhe, Germany\\
$^c$ Department of Physics, University of Oradea, Oradea, Romania\\
$^d$ Research Institute for Solid State Physics and Optics,
P.O.\ Box 49, H-1525 Budapest, Hungary
}

\date{\today}
\begin{abstract}
We generalize the spectral sum rule preserving density matrix numerical
renormalization group (DM-NRG) method in such a way that it can make use of  
an arbitrary number of not necessarily Abelian, local symmetries present 
in the quantum impurity system. We illustrate the benefits of using
non-Abelian symmetries by  
the example of calculations for the $T$-matrix of the two-channel Kondo model 
in the presence of magnetic field, for which conventional NRG methods
produce large errors and/or take a long run-time. 
\end{abstract}
\maketitle

\section{Introduction}
Quantum impurity models play a crucial role in our understanding of strongly
correlated systems: They appear in the description of correlated mesoscopic
structures,\cite{QD_review} they show up in molecular electronics, and many of the
properties of correlated bulk systems can also be accounted for using
self-consistent quantum impurity  models within the dynamical mean field
approach.\cite{Georges}   
Despite the lot of interest and the large amount of effort invested in
understanding these models, we have, unfortunately, very limited tools to
describe quantitatively the general properties of a generic quantum impurity
model. For some of the quantum impurity models Bethe
Ansatz,\cite{Andrei_84,Tsvelick_85} conformal field theory,\cite{Affleck_Ludwig_91} bosonization,\cite{Emery_Kivelson} 
perturbative calculations,\cite{Abrikosov,Fowler} or a Fermi liquid theory\cite{Nozieres_74} can provide a satisfactory  
explanation. However, these results are usually restricted to some
regions in the parameter space. Therefore, even today, the most reliable method to obtain
accurate information on a generic quantum impurity model over the whole parameter
space and for any frequency is Wilson's numerical renormalization group 
method (NRG)\cite{NRG_ref} originally developed for the one-channel Kondo model (1CKM). 

Apart from being extended to compute dynamical
properties,\cite{Oliveira,Costi_92,Costi_94,Suzuki} Wilson's 
method has been used in its original form for a longtime, and it is 
only recently that this method has been further developed. First, Hofstetter
realized that a density matrix NRG (DM-NRG) procedure needs to be introduced in
certain cases to avoid spurious results.\cite{Hofstetter_00} The method of Hofstetter, however, 
applied the original truncation scheme of Wilson, and therefore has not conserved spectral weights.
A remarkable development of the NRG scheme has been the introduction of 
a complete basis set by  Anders et al.,\cite{Anders_05} which was used to
develop the time-dependent DM-NRG algorithm, and it also led to the development of
spectral sum-conserving DM-NRG algorithms.\cite{Peters_06,Weichselbaum_07} 
In the work of Weichselbaum and von Delft, the spectral sum conserving method
has been hybridized with a matrix product state approach.\cite{Weichselbaum_07} 

Using symmetries is an important element of NRG. 
For the simplest models, it is usually sufficient to use Abelian symmetries.
However, for the more interesting two- and multi-channel models it is crucial
to  exploit symmetries, since the computational effort needed increases
rapidly with the number of electron channels, and it is very important to keep
a sufficiently large number of levels in the DM-NRG procedure to achieve good accuracy.
Despite the amazing development discussed in the previous paragraph, a
general framework in which the advantages of  non-Abelian 
symmetries are exploited in conjunction with DM-NRG was still
missing, apart from the generalization of DM-NRG for the case of only one
SU(2) symmetry.\cite{Zitko_06} 
The aim of the present paper is to develop such a scheme and to
show, how the spectral sum-conserving DM-NRG method can be used 
in combination with  non-Abelian symmetries. For this purpose, we derive a very general
recursion formula for the reduced density matrix. We shall show how this
formula can be used to evaluate the spectral function of any retarded Green's
function  of the form
\be 
G^{R}_{A,B^\dagger}(t) \equiv - i 
\left\langle [{ A}(t),{ B}(0)
]_\xi \right\rangle \Theta(t)\;.
\label{G_R}
\ee
Here $A$ and $B$ are any kind of local fermionic 
or bosonic operators acting at the
impurity site, and  $\left[ A, B\right]_{\xi}= AB-\xi BA$ denotes the
commutator ($\xi=1$) and the anticommutator ($\xi=-1$) for bosonic 
or fermionic operators, respectively. 
In Eq.~\eqref{G_R},  $\langle\dots\rangle$ denotes  the average
with the equilibrium density matrix,  $\langle\dots \rangle=
\textrm{Tr}\left\{ \dots {\varrho} \right\}$. 
Expressions for static quantities shall also be derived.

Due to the general recursion relation mentioned above, we were able to
perform a DM-NRG  calculation independently of the 
type and number  of symmetries considered, and to build a
completely flexible DM-NRG code that handles symmetries dynamically and 
``blindly''.\cite{future_cite}

To demonstrate the benefits of our procedure, we
shall present numerical results for the two-channel Kondo model,
which is the most basic non-Fermi liquid
quantum impurity model,\cite{Cox} and provides an excellent testing ground for 
multiple non-Abelian symmetries. Furthermore, its study is also 
motivated by its recent mesoscopic realization in double-dot systems by Potok 
{\sl et al.}\cite{Potok_07} As we shall see, conventional NRG techniques and 
DM-NRG with Abelian symmetries give rather poor results compared to a 
computation performed using the non-trivial symmetry structure of the model.

The paper is organized as follows. In Section \ref{sec:sym} we present how
one can exploit the internal symmetries of the quantum impurity system 
in the NRG process. In Section \ref{sec:construction} the general procedure, which permits 
the use of an arbitrary number of local
symmetries, is extended to the spectral sum rule preserving DM-NRG
algorithm. In Section \ref{sec:spectral} we provide the
general formulas for computing Green's functions in the DM-NRG framework using symmetries. 
In Section \ref{sec:numres} we present our numerical results for the local
fermions' and local composite fermions' spectral functions and for the on-shell 
$T$-matrix in the two-channel Kondo model which strongly support the use of 
DM-NRG (with the largest possible symmetry of the system) as opposed to the use
of NRG.   In Section \ref{sec:conclusion} we sum up what has been presented.
In Appendix \ref{app:SU2} we give an alternative derivation of the result 
that the reduced density matrix retains its diagonal form in the course of the
reduction for the special case when the symmetry of the system contains only
SU(2) groups from among non-Abelian groups.

\section{The role of symmetries in the NRG procedure}\label{sec:sym}
In the present section we shall briefly discuss, how non-Abelian symmetries
appear in the  NRG calculations. However, before setting up the
general formalism it is useful to discuss the role of non-Abelian
symmetries in the specific example of the one-channel Kondo (1CK) model 
and consider the general case and the general recursion relations only afterwards.

\subsection{Symmetries of the one-channel Kondo model}

In the NRG procedure, Wilson used the following approximation 
for the 1CK Hamiltonian\cite{NRG_ref}   
\begin{multline}
H_{1CK} = \frac{1}{2}\; {\cal J}\sum_{\sigma, \sigma' \in \{\uparrow, \downarrow \}} 
{\vec S} f_{0, \sigma}^{\dagger}{\vec \sigma}_{\sigma\sigma'}^{}f_{0,\sigma'}^{} \\
+\sum_{n=0}^{\infty}\;\sum_{\sigma\in \{\uparrow, \downarrow\}}
t^{}_n \left ( f_{n,\sigma}^{\dagger} f_{n+1,\sigma}^{} + h.c.  \right) \,.
\label{eq:1_chan_Kondo}
\end{multline}
In this Hamiltonian $f^\dagger_{0,\sigma}$ creates a conduction electron of
spin $\sigma$ at the impurity site, and thus the first term 
describes the interaction  between  
the impurity spin, ${\vec S}$, and this  local fermion   through an exchange coupling, 
${\cal J}$. The dynamics of this local fermion is  described by a
semi-infinite chain, the Wilson chain. Electrons (fermions) move along this semi-infinite
chain with an exponentially decreasing hopping amplitude, 
$t_n\propto\Lambda^{-n/2}$, with $\Lambda$ the discretization parameter.\cite{NRG_ref}

The above model has  an $\textrm{SU}_{S}(2)$ symmetry, corresponding to 
spin rotations, i.e. it is invariant under unitary transformations
\be 
H_{1CK} = {\cal U}_s\; 
H_{1CK}\;{\cal U}_s^\dagger \;,
\ee
where the unitary operator ${\cal U}_s $ is generated by the total
spin operators, 
\bea
{\cal U}_s &\equiv& {\cal U}_s ({\vec \omega}_s) = e^{i {\vec \omega}_s \; {\vec S}_{T} } \;,
\\
{\vec S}_{T} &=& {\vec S} +\frac{1}{2}\sum_{n=0}^{\infty} \sum_{\sigma, \sigma' \in\{\uparrow ,\downarrow \}}
f_{n,\sigma}^{\dagger} {\vec \sigma}_{\sigma, \sigma'}^{} f_{n,\sigma'}^{}\;.\label{eq:generators_spin}
\eea

The Hamiltonian $H_{1CK}$ is also invariant under the action of  $\textrm{SU}_{C}(2)$
rotations  in charge  space, ${\cal U}_c  = e^{i {\vec \omega}_c \; {\vec C} } $,
generated by the operators $C^{x}=(C^{+}+C^{-})/2$, $C^{y}=(C^{+}-C^{-})/2i$ and $C^z$ with 
\begin{eqnarray}\label{eq:generators_charge}
C^{+} & = & \sum_{n=0}^{\infty} (-1)^n f_{n,\uparrow}^{\dagger}f_{n,\downarrow}^{\dagger}\;, \nonumber\\
C^{z} & = & \frac{1}{2}\sum_{n=0}^{\infty} \sum_{\mu=\{\uparrow, \downarrow \}} (f_{n, \mu}^{\dagger}f_{n,\mu}-1)\;,\\
C^{-} & = & \left(C^{+}\right)^{\dagger}\,.\nonumber
\end{eqnarray}

Since the spin symmetry generators commute with the charge symmetry
generators, $H_{1CK}$ possesses a 
symmetry,  $\textrm{SU}_{S}(2)\times\textrm{SU}_{C}(2)$,\cite{Jones_87} 
and consequently, 
the eigenstates of the Hamiltonian form degenerate multiplets, 
$|i,{\underline Q}_i, {\underline Q}_i^z\rangle$, that are classified by their 
label $i$, the spin and charge 
quantum numbers: ${\underline Q}_i \equiv\{ S_i, C_i \}$,\footnote{The eigenvalues
of the operators ${\vec S}^2$ and ${\vec C}^2$ are given by $S(S+1)$, $C(C+1)$.}
and the $z$-components of the charge and spin operators, ${\underline Q}^{z}_i \equiv\{ S^{z}_i,C^{z}_i \}$, 

In the following, we shall refer to the quantum numbers 
${\underline Q} $ as {\em  representation indices},   
while  ${\underline Q}^{z}$ are referred to as {\em labels} of the internal states of a given multiplet.
Note that the representation index ${\underline Q}$ defines the dimension of the
given irreducible subspace which is ${\rm dim}(i) = {\rm dim}(\underline
Q_i)= (2S_i+1)(2C_i+1)$ in the example above.

Similar to irreducible subspaces of the Hilbert space, operators can be organized into irreducible
tensor operator multiplets.\cite{Cornwell} One of the simplest examples is
provided by the impurity spin, from the components of which we 
can form the operator triplet as  
\be
\{A_m\} \equiv \left\{ -\frac{1}{\sqrt{2}}\; S^{+}, S^z, \frac{1}{\sqrt{2}}\;S^{-} \right\}\;.
\ee 
The components of this triplet transform  under spin rotations 
as the eigenstates $|m\rangle$ of $S^z$ within a spin
$S=1$ multiplet, while they are invariant under charge
rotations. This means  that $A$ has quantum numbers, $S_A = 1$ and $C_A=0$,
and the components of this operator multiplet are labeled by
$S^z_A=m, (m=0,\pm1)$, while the internal charge label is trivially
$C^z_A=0$. Similar to the eigenstates of the Hamiltonian, 
 in our simple example the quantum numbers of an   
irreducible tensor operator $B$ can be organized into a representation  
index vector, ${\underline b} = (S_B,C_B)$ and its components are labeled by 
${\underline b}^z = (S^z_B,C^z_B)$ taking the values $S^z_B = -S_B,-S_B+1,\dots,S_B$
and $C^z_B = -C_B,-C_B+1,\dots,C_B$. A further example for an $S=1/2$ and
$C=1/2$ operator is formed 
by the four operators $\{f^\dagger_{0,\uparrow},f^\dagger_{0,\downarrow}, -f_{0,\downarrow}, f_{0,\uparrow}\}$.

\begin{widetext}
The Wigner--Eckart theorem\footnote{Some of the considerations
  presented here do not carry over for non-compact Lie Groups.} then tells us that, apart from trivial group
theoretical factors (Clebsches), the matrix elements of the members of a given
operator multiplet and states within two multiplets, $i$ and $j$ are simply
related by  
\be
\left < i,{\underline Q}_{i}  \underline Q^z_{i} \right|  B_{\underline b, \underline b_z}
 \left | j , \underline Q _{j} \underline {Q}^z_{j}\right > = 
\left < i\right . \parallel B\parallel \left . j  \right > 
\left < \underline Q_{i}  \underline Q^z_{i} \right | \underline b, \underline b_z; 
\left .  \underline Q _{j} \underline {Q}^z_{j}\right >\;
\label{eq:Wigner_Eckart}
\ee
where $\left < i\right . \parallel B\parallel \left . j  \right > $ denotes the
reduced (invariant) matrix element of $B$, and the generalized Clebsch--Gordan coefficients are simply defined as 
\be
\left < \underline Q_{i}  \underline Q^z_{i} \right | \underline b, \underline b_z; 
\left .  \underline Q _{j} \underline {Q}^z_{j}\right >
= 
\left < S_{i} S^z_{i} \right | S^{}_B,S^z_B;  \left . S _{j}  {S}^z_{j}\right >
\;
\left < C_{i} C^z_{i} \right | C^{}_B,C^z_B;  \left . C _{j}  {C}^z_{j}\right >\;,
\ee
with the usual SU(2) Clebsches\cite{Cornwell} on the right hand side. 
This relation is used extensively in the NRG calculations. 
\end{widetext}

An important property of the unitary transformations above is that they are
{\em local} in the sense that they decompose into unitary
operators which commute with each other and act independently at different sites, 
\bea 
{\cal U} &=& {\cal U}_s \times {\cal U}_c\;,
\\
{\cal U}_c &=& \prod_n  {\cal U}_{c,n}\;,
\\ 
{\cal U}_s &=& \prod_n  {\cal U}_{s,n}\;.
\eea
This decomposition property is crucial for using symmetries in the  NRG
calculations.

\subsection{General local symmetries on the Wilson chain}

Under rather general conditions, the symmetry considerations above and the NRG
procedure  can be extended to Hamiltonians of the  form 
\be
\label{eq:chain_hamiltonian} 
H = {\cal H}_0 + \sum_{n=0}^{\infty} \left( {\tau}_{n,n+1} + {\cal H}_{n+1}\right )\;.
\ee 
Here ${\cal H}_0$ contains the interaction between the impurity and the
fermionic bath (the site of the impurity is labeled by 0), and
nearest-neighbors  on the Wilson chain are coupled through the
hopping terms, ${\tau}_{n,n+1}$. The $n^{th}$ on-site term  
 ${\cal H}_{n+1}$ describes  local correlations/interactions, 
and it can also account for the absence of the electron-hole symmetry.

\begin{figure}[b]
\includegraphics[width=0.7\columnwidth,clip]{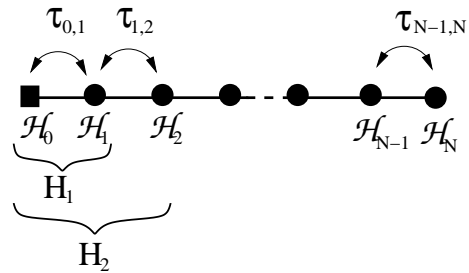}
\caption{ Construction of the Hamiltonian on the Wilson chain of length $N$. 
The impurity sits at the square shaped site labeled by 0, dots represent
the further sites. 
At the zero$^{th}$ NRG iteration ${H}_0$ is identical with
${\cal H}_0$. 
As the $n^{th}$ site is added to the chain, $H_n$ 
is constructed recursively from the hopping term (${\tau}_{n,n+1}$),
from the on-site energy (${\cal H}_n$), and from $H_{n-1}$.
}
\label{fig:wilson_chain}
\end{figure}

The usual NRG solves the model  Eq.\ (\ref{eq:chain_hamiltonian})
by an iterative diagonalization process. The iteration steps consist
of diagonalizing the set of Hamiltonians introduced recursively by  
\bea
H_{0}&=&{\cal H}_0\;,\\ 
H_{n+1}& =& H_n +\tau_{n,n+1}+{\cal H}_{n+1}\,.
\label{eq:recursive_hamiltonian}
\eea
This recursion is depicted  in Fig.\ \ref{fig:wilson_chain}. 

Let us now assume that $H$ as well as every $H_n$ is invariant
under the group $G$, i.e.\
\bea
{\cal U}(g)\;H_n\; {\cal U}^{-1}(g)= H_n\,,\quad\quad n=0,1,2,\dots\;,
\eea
holds for every $g\in G$, with ${\cal U}(g)$ the appropriate  
unitary operator.  Furthermore, let us suppose that $G$ and correspondingly 
${\cal U}$ can be decomposed 
into a direct product of $\Gamma$ subgroups 
 ${\cal  G}_{\gamma}$ $(\gamma = 1,...,\Gamma)$, each acting independently on every
lattice site, 
\bea
G &=&  {\cal G}_{1}\times{\cal G}_{2}\times\dots\times{\cal G}_{\Gamma}\,, 
\\
{\cal U}(g) &=& \prod_{\gamma=1}^{\Gamma} {\cal U}_\gamma (g_{\gamma}) =
\prod_{\gamma=1}^{\Gamma}\prod_n {\cal U}_{\gamma,n} (g_{\gamma}) 
\,, 
\eea
Note that in the 
considerations above the subgroups can be also finite, and $G_\gamma$ can represent
a crystal field symmetry as well as e.g.\ the SU(3) group. However, some of the 
considerations presented in this paper may not apply for non-compact groups.

The above decomposition is not necessarily unique. 
Nevertheless, having obtained a specific decomposition of the symmetry, 
the argument of the previous  subsection can be repeated 
with the only difference that now $\Gamma$ number of quantum numbers classify the
irreducible subspaces (multiplets)  of the Hamiltonians $H_n$,
\be
\underline{Q} = \left\{Q^1, Q^2,...,Q^{\Gamma}\right\}\;
\ee
and states within the multiplet are then labeled by the internal quantum numbers 

\be
\underline {Q}^z =\left\{ Q^{1,z}, Q^{2,z},...,Q^{\Gamma,z} \right\}\;. 
\ee
Similar to the simple 1CK example, the dimension of a subspace $i$ depends uniquely 
on its quantum numbers $\underline{Q}_i$, i.e.\ ${\rm dim}(i)={\rm dim}(\underline Q_i)$. 

Operators can also be arranged into irreducible tensor
operators, and an irreducible tensor operator multiplet $A$ is correspondingly
described  by quantum numbers $\underline a$, while members of the multiplet are labeled 
by $\underline a^z$ with $\underline a$ and $\underline a^z$ being
$\Gamma$-component vectors. The Wigner--Eckart theorem,
Eq.~\eqref{eq:Wigner_Eckart}, carries over to the general case too with the
slight modification that the Clebsch--Gordan coefficients are now defined as 
\begin{multline}
\left<\underline{Q}^{}_i\;\underline{Q}^z_i\;\left|\underline{a}\;\underline{a}^{z};\underline{Q}^{}_j\;
\underline{Q}^z_j\right>\right.
\nonumber \\
\equiv \prod_{\gamma=1}^{\Gamma}\left<Q^{\gamma}_i\;Q^{\gamma,z}_{i}\;
\left|a^{\gamma}\;a^{\gamma,z}\;Q^{\gamma}_j;Q^{\gamma,z}_j\right>\right.\;.
\label{eq:general_Clebsch}
\end{multline}

\subsection{The role of symmetries in the diagonalization procedure}

Before discussing NRG with a  complete basis set,\cite{Anders_05} let us
shortly look into how 
symmetries are used in Wilson's NRG. In his original work, Wilson constructed approximate eigenstates of $H_n$ for a chain of length $n$,
iteratively. However, in each iteration the dimension of the Hilbert space
increases by a factor $d$, with $d$ the dimension of the local Hilbert space
at a single site of the chain with site label $n>0$. Therefore the size of the Hilbert space increases
exponentially with $n$, and after a few iterations one must truncate it: 
 Some of the states $i$ are therefore discarded ($i\in D$),
while other states are kept ($i\in K$), and are used to construct 
approximate eigenstates for $H_{n+1}$. 
Symmetries are of great value in this diagonalization procedure: 
In their presence the $H_n$'s are block diagonal in the representation
indices, and the eigenvalue problem can be solved much more efficiently. 
For some of the physical quantities, it is crucial to increase the number of
kept states as much as possible  to achieve good
numerical accuracy.

In Wilson's original formulation of NRG, symmetries are used in the following way: 
As discussed above,  in the $n^{th}$ iteration the eigenstates (multiplets) of $H_n$ are constructed
from the kept multiplets labeled by $u$ with $u\in K$ of the $(n-1)^{th}$
iteration, which are approximate low-energy eigenstates of
$H_{n-1}$, and from a complete set of  \emph{local states} (multiplets),
labeled by $\mu$, that live at the $n^{th}$ site.  In the following, we refer to these new approximate eigenstates  $i$ 
as \emph{new states}, while by analogy with DMRG, we shall call the kept states
\emph{block states} or \emph{old states}. 

In the presence of symmetries, each 
new multiplet carries representation indices $\underline {Q}_{\tilde i}= \{
Q_{\tilde i}^{\gamma} \}$, 
and states within this multiplet are labeled by the 
 internal quantum numbers $\underline {Q}_{\tilde i}^z= \{Q_{\tilde i}^{\gamma,z}
\}$. Similarly,  local states have quantum numbers 
$\underline {q}_{\mu}= \{ q_{\mu}^{\gamma} \}$
and are labeled by $\underline {q}_{\mu}^z= \{ q_{\mu}^{\gamma,z} \}$. 
To construct the approximate eigenstates of $H_n$, we first use the Clebsches
to build new states from the block (old) and local states that transform as irreducible 
multiplets under the symmetry transformations, ${\cal U}(g)$, 
\begin{widetext}
\be
\left | \tilde i,  \underline{Q}_{\tilde i}, \underline {Q}_{\tilde i}^z \right >_{n} \equiv 
\sum_{\underline {Q}_u^z, \underline {q}_{\mu}^z}
\left < \underline{q}_{\mu} \underline{q}_{\mu}^z; \underline{Q}_u \underline{Q}_u^z \right | 
\left . \underline{Q}_{\tilde i} \underline {Q}_{\tilde i}^z  \right > 
\left | u, \underline{Q}_{u}, \underline{Q}_u^z; \mu, \underline{q}_{\mu}, \underline{q}_{\mu}^z \right >_{n-1}\,, 
\phantom{nnnn}(u\in {\rm K})  
\label{eq:new_states}
\ee
\end{widetext}
These states shall be referred to  as the  
{\it canonical basis} from now on. In this basis
 $H_n$ has a block-diagonal structure
\begin{multline}
\Big._{n}\left< \tilde i,\;  \underline{Q}_{\tilde i},\; \underline
{Q}_{\tilde i}^z\; 
\left|\;H_n\;\right|\; \tilde j,\;  \underline{Q}_{\tilde j},\; \underline {Q}_{\tilde j}^z
\right>_{n}
\nonumber
\\
= 
\big._n\left< \tilde i \parallel H_n\parallel \tilde j\right>_n\;
\delta_{\underline{Q}_{\tilde i},\underline{Q}_{\tilde j}} 
\delta_{\underline{Q}^z_{\tilde i},\underline{Q}^z_{\tilde j}} 
\;, 
\end{multline}
and is diagonalized by a unitary transformation, ${\cal O}^{[n]}_{\tilde i, i }$ 
\bea
&\left | i,  \underline{Q}_{i}, \underline {Q}_{i}^z \right >_{n} = 
\sum_{\tilde i} {\cal O}^{[n]}_{\tilde i, i } 
\left | \tilde i,  \underline{Q}_{\tilde i}, \underline {Q}_{\tilde i}^z \right >_{n}
\delta_{\underline{Q}_{i}, \underline{Q}_{\tilde i}}\delta_{\underline {Q}_{i}^z, \underline {Q}_{\tilde i}^z}\,.
\label{eq:rotated_new_states}
\\
& H_n \left|i,  \underline{Q}_i,\underline{Q}^z_{i}\right>_{n}= 
 E^n_{i} \left|i,\underline{Q}_i,\underline{Q}^z_{i}\right>_{n}\,,
\label{eq:eigenvalues}
\eea
with $E^n_{i}$  the eigenenergies of $H_n$. 
Here ${\cal O}$ is a block-diagonal matrix, and its  columns in a given symmetry
sector are just the eigenvectors 
of the corresponding submatrix of  $\big . _n \langle \tilde i\parallel H_{n}\parallel\tilde
j\rangle \big . _n$ in the canonical basis.  
In the upcoming iteration, some of these
multiplets shall be kept, and form the block states for $(n+1)^{th}$ iteration, while others are again discarded.   

In the iteration step outlined above we need the matrix elements 
$\big ._n \left <\tilde i \right . \parallel H_n\parallel\left. \tilde j   \right>\big . _n $, 
with $H_n = H_{n-1}+\tau_{n-1, n}+{\cal H}_n$.
The matrix elements of $H_{n-1}$ are simply
\begin{equation}
\big. _n \left < \tilde i \right . \parallel H_{n-1} \parallel\left.\tilde j  \right > \big . _n= 
E^{n-1}_{u}\; \delta_{\tilde i, \tilde j}\;, 
\end{equation}
where $u$ is the state from which state $\tilde i$ has been constructed. 
Similarly, matrix elements of ${\cal H}_n$
are given by
\begin{equation}
\big. _n \left < \tilde i \right . \parallel {\cal H}_n \parallel\left.\tilde j  \right > \big . _n= 
\varepsilon^{n}_{\mu}\; \delta_{\tilde i, \tilde j}\; , 
\end{equation}
where $ \varepsilon^{n}_{\mu}$ is just the expectation value of ${\cal H}_n$ with local states within 
the multiplet $\mu$. 
Finally, to compute the matrix elements of the hopping  $\tau_{n-1, n}$, 
we use the fact that by symmetry, $\tau_{n-1, n}$ can always be decomposed as
\begin{equation}
\tau_{n-1, n} =\sum_{\alpha}h_{\alpha }^{[n-1]}\sum_{{\underline c}^z}\left [C^{[n-1]}_{\alpha; \underline c, \underline c^z}
\left (C^{[n]}_{\alpha; \underline c, \underline c^z}\right )^\dagger + H. c. \right ]\; . \label{eq:hopping_term}
\end{equation}
Here $C^{[n-1]}_{\alpha; \underline c, \underline c^z}$ denotes some {\em creation operator multiplet} at site 
$n-1$ that has quantum numbers $\underline c$,  and $h_{\alpha }^{[n-1]}$'s are the  hopping 
amplitudes between sites $n-1$ and $n$. The index $\alpha$ in the equation above labels
various `` hopping operators``. To give a simple example, if we treat the 1CKM using only $U(1)$
symmetries then  one has two hopping operators,  $\alpha \in \{1,2\}$, corresponding to 
$C_1^{[n]} = f^\dagger_{n, \uparrow}$ 
and $C_2^{[n]} = f^\dagger_{n, \downarrow}$. However, if we use the spin $SU(2)$ symmetry, then $\alpha =1$ and 
 $C_1^{[n]} = \left \{f^\dagger_{n, \uparrow},  f^\dagger_{n, \uparrow}   \right \}$. 

For the reduced matrix elements of $\tau_{n-1, n}$ one obtains using Eq. \eqref{eq:new_states}
and the decomposition,  Eq.\ \eqref{eq:hopping_term}, the following formula,  
\begin{widetext}
\begin{multline}
\big._{n}\left<\tilde i\parallel\tau^{}_{n-1,n}\parallel
\tilde j\right>\big._{n}=\ds{\sum_\alpha}\;
\big._{n-1}\left<u\parallel C^{[n-1]}_\alpha\parallel v\right>\big._{n-1}\; 
\left<\nu\parallel C^{[n]}_\alpha\parallel
\mu\right>^\ast\;\times\\
 D\left(\alpha, \underline c, \underline Q_{\tilde i}, \underline Q_{\tilde j}, 
\underline Q_u, \underline Q_v,\underline q_{\mu},{\underline q}_\nu \right)
\delta_{\underline{Q}_{\tilde i},\underline{Q}_{\tilde j}}
\delta_{\underline{Q}^z_{\tilde i},\underline{Q}^z_{\tilde j}}
\;.
\end{multline}
\end{widetext}
Here the state $\tilde i$ has been constructed from the state $u$ of the
previous iteration and from the local state $\mu$, 
while $\tilde j$ has been constructed from $v$ and $\nu$. 
The functions
 $ D\left(\alpha,\underline c, \underline Q_{\tilde i}, \underline Q_{\tilde j}, 
\underline Q_u, \underline Q_v,\underline q_{\mu},{\underline q}_\nu \right) $ denote 
 group theoretical factors, 
 \begin{widetext}
\begin{multline}
 D\left(\alpha,\underline c, \underline Q_{\tilde i}, \underline Q_{\tilde j}, 
\underline Q_u, \underline Q_v,\underline q_{\mu},{\underline q}_\nu \right)=
\textrm{sgn}(C,\;\mu)\;
\ds{\sum_{\underline{C}^z}}
\ds{\sum_{\underline{Q}^z_u,\;\underline{Q}^z_v}}\;
\ds{\sum_{\underline{q}^z_\mu,\;\underline{q}^z_\nu}}
\left < \underline Q_{\tilde i}  \underline Q^z_{\tilde i} \right | \underline{q}_\mu\, \underline{q}^z_\mu; 
\left .  \underline Q_{u} \underline {Q}^z_{u}\right >
\left < \underline Q_{\tilde i}  \underline Q^z_{\tilde i} \right | \underline{q}_\nu\, \underline{q}^z_\nu; 
\left .  \underline Q_{v} \underline {Q}^z_{v}\right >^*\times \\
\left < \underline Q_{u}  \underline Q^z_{u} \right | \underline c\, \underline c^z; 
\left .  \underline Q_{v} \underline {Q}^z_{v}\right >
\left < \underline q_{\nu}  \underline q^z_{\nu} \right | \underline c\, \underline c^z; 
\left .  \underline q_{\mu} \underline {q}^z_{\mu}\right >^*\;,
\end{multline}
\end{widetext}
where the sign function $\textrm{sgn}(C,\mu)=\pm 1$ arises as one commutes the creation 
operators implicitly present in the local state $\mu$ over the operator 
$C^{[n-1]}$, and it is negative if $C^{[n]}_{\alpha}$ is a 
fermionic operator and  the local state  $\mu$  is constructed from an odd number
of fermions, otherwise it is positive. 
The local matrix elements, $\langle \mu\parallel  C^{[n]}_\alpha \parallel \nu\rangle$ 
are the same for all sites and can easily be
determined, while $\big . _{n-1}\langle u\parallel C^{[n-1]}\parallel v\rangle\big . _{n-1}$ 
can be computed from the previous iteration by recursion. 
In fact, for any operator $A$, acting on sites $m<n$, 
and  whose matrix elements
are known in the iteration $n-1$, we have the following recursion relation
\begin{widetext}
\begin{equation}
\big ._n \left < \tilde i\right .\parallel A \parallel \left . \tilde j\right > \big ._n = 
\big ._{n-1}\left < u\right .\parallel A \parallel \left . v \right > \big ._{n-1}\;
 F\left (\underline a,  \underline Q_{\tilde i}, \underline Q_{\tilde j}, 
\underline Q_u, \underline Q_v,\underline q_{\mu} \right )\;\delta_{\underline q_\mu, \underline q_\nu}\;
,
\end{equation}
where the factor $ F\left (\underline a,  \underline Q_{\tilde i}, \underline Q_{\tilde j}, 
\underline Q_u, \underline Q_v,\underline q_{\mu} \right )$ 
is given by the following expression
\begin{multline}
 F\left (\underline a,  \underline Q_{\tilde i}, \underline Q_{\tilde j}, 
\underline Q_u, \underline Q_v,\underline q_{\mu} \right ) = 
\textrm{sgn}(A,\;\mu)\;
\frac 1{{\rm dim}(a) {\rm dim}(\tilde j)}\sum_{\underline Q_{\tilde j}^z,\underline a^z }
{\left< \underline Q_{\tilde i}\, \underline Q_{\tilde i}^z  \right .\left | 
\underline a\, \underline a^z ;  \underline Q_{\tilde j}\,  \underline Q_{\tilde j}^z \right >^* }   
\times \\
\sum_{ \underline Q_u^z,  \underline Q_{v}^z }\sum_{ \underline q_{\mu}^z} \;
\left< \underline Q_{\tilde i}\, \underline Q_{\tilde i}^z  \right .\left | 
\underline q_\mu\, \underline q_\mu^z ;  \underline Q_{u}\,  \underline Q_{u}^z \right > 
\left< \underline Q_{\tilde j}\, \underline Q_{\tilde j}^z  \right .\left | 
\underline q_\mu\, \underline q_\mu^z ;  \underline Q_{v}\,  \underline Q_{v}^z \right > ^*
\left< \underline Q_{u}\, \underline Q_{u}^z  \right .\left | 
\underline a\, \underline a^z ;  \underline Q_{v}\,  \underline Q_{v}^z \right > 
\end{multline}
\end{widetext}

\begin{figure}[ht]
  \includegraphics[width=0.9\columnwidth,clip]{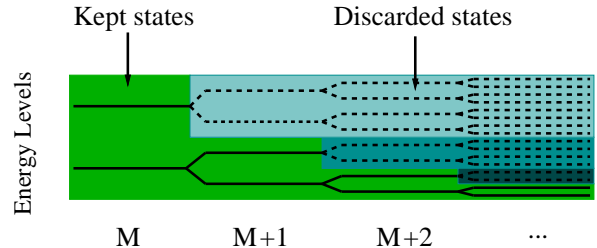}
  \caption{(color online) A complete basis of a Wilson chain 
    represented as the exponentially increasing number of energy levels
    belonging to the successive iterations. Continuous/dashed lines represent kept,
    low-energy/discarded, high-energy levels, respectively. For the consecutive
    iteration steps the distances between the levels illustrates how the
    energy resolution of NRG gets exponentially refined. 
  }
  \label{fig:energy_levels}
\end{figure}

One drawback of the algorithm above is that the 
eigenstates of $H_n$ constructed this way 
do not form a complete basis set on the Wilson chain of length $n$, 
since states descendant from the 
discarded states of the previous iteration are missing. 
However, as it was recently shown,\cite{Anders_05} one can construct a
complete basis of the Wilson chain in a slightly different way: Let us
consider a chain of length $N$, and construct approximate 
eigenstates of $H_n$ that, however, live at {\em all sites} of this chain, 
\be 
\left|i, \underline{Q}_{i}, \underline {Q}_{i}^z \right>_{n}
\rightarrow \left|i, \underline{Q}_{i}, \underline {Q}_{i}^z; e \right>_{n}\;.
\ee
In this equation $e$ just labels the $d^{N-n}$ independent 'environment' states living at the last 
$N-n$ sites of the chain. The internal structure of these environment states 
is not important, only their degeneracy shall play some role. The previous
iterative construction carries over to these states, too.  
By construction, discarded  states (together with their environment state) 
form a complete basis set:
\be
\1= \sum_{n=0}^{N} \sum_{i\in {\rm D}}\sum_{e}
\sum_{\underline{Q}^z_i}
\left|i, \underline{Q}_i,\underline{Q}^z_{i};e\right>_{n}\Big._{n}
\left<i,\underline{Q}_i,\underline{Q}^z_{i};e\right|\,,
\label{eq:completeness}
\ee
where $i\in D$ refers to the fact that \emph{only discarded states} appear in the sum. 
In Eq. \eqref{eq:completeness}  all states are considered
discarded in the last iteration, $n=N$. We remark that, in reality, the summation in the expression above 
starts only at a value $n=M$, where the first truncation is carried out.
Fig.\ \ref{fig:energy_levels} illustrates the structure of this complete basis.  
In the formulation of the DM-NRG algorithm we shall use
 several times the completeness relation,  Eq.\ (\ref{eq:completeness}).

\section{Construction of the reduced density matrix using symmetries}\label{sec:construction}

In the DM-NRG procedure on a Wilson chain of length $N$, 
the equilibrium density matrix is approximated by  

\begin{eqnarray}
\varrho &=&\ds \sum_{n=0}^{N}\; \varrho^{[n]}\\
\varrho^{[n]} &=& {\sum_{\underline{Q}^z_i,i, e}}
\frac{e^{-\beta E_i^n}}{\cal Z}
\left|i, \underline{Q}_i,\underline{Q}^z_{i}; e\right>_{n}\Big._{n}
\left<e; i,\underline{Q}_i,\underline{Q}^z_{i}\right|\;,\nonumber\\
\label{eq:DM}
\end{eqnarray}
with $\beta =1/k_B T$ the Boltzmann factor and
\be
{\cal Z} =\ds{\sum_{n=0}^{N}}\, \ds{\sum_{{\underline{Q}^z_i,i}}}
e^{-\beta E_i^n} d^{N-n}\,
\label{eq:partition_function}
\ee
the partition function. In Eq.\ (\ref{eq:partition_function}) the factor
$d^{N-n}$ accounts for the degeneracy of the environment states in iteration $n$, 
i.e. for the local degrees of freedom at sites $m>n$.
Since the eigenenergies do not depend on the internal quantum numbers, the
expression for the partition function can be simplified to 
\bea
{\cal Z} =\ds{\sum_{n=0}^{N}}\, \ds{\sum_{i}}\, \dim (i)\; e^{-\beta E_i^n}
d^{N-n}\;,\\ \dim (i)\equiv \ds{\prod_{\gamma=1}^{\Gamma}}\dim (Q_i^{\gamma})\;,
\eea
with $\dim (Q_i^{\gamma})$ being the dimension of an irreducible subspace
having the representation index $Q_i^{\gamma}$.

The concept of the reduced density matrix\cite{Hofstetter_00} arises naturally 
as one starts to calculate Green's functions with NRG. 
More precisely, the quantity that shows up in the calculations is the 
{\em truncated reduced density matrix}, defined as
\begin{equation}
R^{[n]}=\underset{\left\{ e_n \right\}}{\tr} \left\{\sum_{m > n} \varrho^{[m]}   \right\}\; ,
\end{equation}
where one traces over environment states $e_n$ at sites $m>n$. This truncated 
reduced density matrix clearly satisfies the recursion relations
\begin{eqnarray}
R^{[N]} & =& \varrho^{[N]}\nonumber\; , \\
R^{[n-1]} & = & \sum_{i\in D } \frac{e^{-\beta E^{n-1}_i} d^{N-n+1} }{\cal Z}
\left|i, \underline{Q}_i,\underline{Q}^z_{i}\right>_{n}\Big._{n}
\left<i,\underline{Q}_i,\underline{Q}^z_{i}\right|\nonumber\\
& + &  \underset{{\rm site}\;\; n }{\tr} \left\{R^{[n]} \right\}\; .\label{eq:recursive}
\end{eqnarray}
Note that the environment variable is \emph {missing} in the first term 
of the second expression since it has been traced over. The first term accounts
for the contribution of discarded states, while the second term 
has matrix elements between the kept states only. 

To construct the matrix elements of $R^{[n]}$ we first show by induction 
that $R^{[n]}$ is scalar under symmetry operations. This is clearly true for 
the first term in Eq.\ (\ref{eq:recursive}). To show that the second term is also 
invariant, we simply need to use the locality property of the symmetry transformations,
i.e.\ that on the first $n$ sites $U(g) \equiv L(g) V(g)$, where $L(g)$ 
transforms the local states on site $n$, while $V(g)$ transforms states at sites 
$0\leq m\leq n-1$, and $L$ and $V$ commute with each other. 
Therefore, because of the trace of operators being invariant under cyclic permutations, we have
\begin{multline}
\underset{{\rm site}\; n}{\tr}\left\{ L \, V\,R^{[n]}\, V^+\, L^+  \right\}=
\underset{{\rm site}\; n}{\tr}\left\{  V\,R^{[n]}\, V^+  \right\}\\
=V\, \underset{{\rm site}\; n}{\tr}\left\{  R^{[n]} \right\}V^+ = 
U\, \underset{{\rm site}\; n}{\tr}\left\{  R^{[n]} \right\}U^+ \; .
\end{multline}
However, since $U\, R^{[n]}\, U^+ = R^{[n]}$ by assumption, we immediately obtain that 
\begin{equation}
\underset{{\rm site}\; n}{\tr}\left\{R^{[n]} \right\} = U \, 
\underset{{\rm site}\; n}{\tr}\left\{ R^{[n]}  \right\}U^+\;\label{eq:trace} , 
\end{equation}
implying that 
\begin{equation}
U\, R^{[n-1]}\, U^+ = R^{[n-1]}
\end{equation}
for $R^{[n-1]}$, too. This equation means that $R^{[n]}$ is  {\em scalar} and 
therefore, by the Wigner--Eckart theorem we have 
\begin{multline}
\Big._n\left<i,\underline{Q}_i,\underline{Q}^z_{i}\left|\,R^{[n]}\,
\right|j, \underline{Q}_j,\underline{Q}^z_{j}\right>\Big._n\\
= \big._n\left <i\right . \parallel R^{[n]} \parallel\left . j\right >\big._n
\delta_{\underline{Q}_i, \underline{Q}_j}\; 
\delta_{\underline{Q}_i^z, \underline{Q}_j^z}\; 
\end{multline}
The matrix elements of $\left <i\right . \parallel R^{[n]} \parallel\left . j\right > $ 
between discarded states simply derive from the first term in Eq. 
(\ref{eq:recursive}).
To perform the trace in Eq. (\ref{eq:recursive}) and to construct the explicit
relation between the kept matrix elements of $R^{[n-1]}$ and $R^{[n]}$ some
more work is needed. First, we rotate $R^{[n]}$ to the canonical basis
\bea\label{eq:basis_tr}
\big._n\left< \tilde i\right . \parallel {\tilde R}^{[n]}
\parallel \left . \tilde j\right >\big._n =
     {\cal O}^{[n]}_{\tilde i, i}
\; \big._n \left < i\right . \parallel R^{[n]}\parallel \left . j \right >\big._n \left ({\cal
    O}^{-1}\right )^{[n]}_{j \tilde j}\,.
\eea
Then, using the fact that $\underset{{\rm site }\; n }{\tr} \left \{R^{[n]} \right\} $ 
is diagonal in the symmetry quantum numbers and labels, we can trace
over the local states at site $n$ using the recursion relation 
Eq.\ (\ref{eq:new_states}) to obtain the matrix elements between the kept states
$u,v \in K$ as
\begin{multline}
\big._{n-1}\left< u\right . \parallel R^{[n-1]} \parallel \left . v \right >\big._{n-1} \\
=\ds{\widetilde{ \sum_{{\tilde i, \tilde j},{\underline q}_{\mu},\mu}}}\;
\frac{\dim\left(\tilde i\right)}{\dim\left(u\right)}\; \big._n \left< \tilde i\right . \parallel
{\tilde R}^{[n]}\parallel \left . \tilde j \right>\big._n
\delta_{\underline{Q}_{\tilde i}, \underline{Q}_{\tilde j}}\,.
\end{multline}

Here the tilde over the sum indicates that in the summation over $\tilde i$
and $\tilde j$ only those states are considered which have been 
constructed from $u$ ($\tilde i \Leftarrow u $) and $v$ 
($\tilde j \Leftarrow v $) in Eq.\ (\ref{eq:new_states}), respectively. 
This is a very powerful expression, which applies to essentially any type of
symmetry.

\section{Spectral function computation}\label{sec:spectral}

The quantity that describes the linear response of a static 
system to a time-dependent perturbation and 
which is to be calculated in the DM-NRG framework is the retarded Green's
function. 
For two irreducible tensor operators it is defined as: 
\begin{equation}
G^{R}_{A_{\underline a, \underline a_z},B^\dagger_{\underline b, \underline b_z}}(t) = 
-i \left < \left [ A_{\underline a, \underline a_z} (t), 
B^\dagger_{\underline b, \underline b_z}(0) \right ]   \right > \Theta \left( t \right).
\end{equation}
By symmetry, however, this Green's function is non-zero 
only if the spectral operators $A$ and $B$ transform accordingly to the same 
representation i.e. 
\begin{equation}
G^{R}_{A_{\underline a, \underline a_z},B^\dagger_{\underline b, \underline b_z}}(t) 
=G^{R}_{A,B^\dagger} (t)\; \delta_{\underline a, \underline b}\;
\delta _{\underline a_z, \underline b_z}\; , \label{eq:green_function_extended}
\end{equation}
and it is independent of the value 
of $\underline a_z=\underline b_z$.
Note that in this expression the representation index $\underline b$ 
and its labels $\underline b_z$ 
are the quantum numbers that characterize the operator 
$B$ and not $B^{\dagger}$. 

In the reduced density matrix formalism we can 
generalize the procedure outlined in Ref.\ [\onlinecite{Anders_05}]
even in the presence of non-Abelian symmetries to obtain the
following form for the Laplace transform of the Green's function
\begin{widetext}
\begin{multline}
G^{R}_{A,B^{\dagger}}(z)=
\ds{\sum^{N}_{n=0}}\;\sum_{i\in D, K}\;\ds{\sum_{(j,k)\notin(\rm K,\rm K)}}
\;\big . _n\left <i \right .\parallel R^{[n]}\parallel \left . j\right >\big . _n\\
 \times\left[
\frac{\big ._n \left < k \right . \parallel A^{\dagger }_{}\parallel \left . j \right >\big . _n^{*}
\; \big . _n \left < k \right . \parallel B^{\dagger }_{} \parallel \left . i\right>\big . _n}
{z+\frac{1}{2}(E^n_i +E^n_j)-E^n_k}\frac{\dim(k)}{\dim(a)}
-\xi\;\frac{\big . _n \left< j \right . \parallel B^{\dagger }_{} \parallel \left . k \right>\big . _n 
\;\big . _n \left<i \right . \parallel A^{\dagger }_{}\parallel \left . k \right>\big . _n^*}
{ z-\frac{1}{2}(E^n_i +E^n_j)-E^n_k } \frac{\dim (i)}{\dim (a)} \right]\;,
\label{eq:green_function_explicite}
\end{multline}
\end{widetext}
Remarkably, this formula contains exclusively the reduced matrix elements 
and the dimensions of the various multiplets. 
Here the second sum is over all the multiplets $i,j,k$ of the given iteration  
 subject to the restriction 
 that $j,k$ do not belong to kept states at the same time and no summation is
 needed for states within the multiplets. In Eq.\ (\ref{eq:green_function_explicite})
$dim(a)={\prod_{\gamma=1}^{\Gamma}}\dim (a^{\gamma})$ is the dimension of the operator
multiplet $A_{\underline a, \underline a_z}$.
We note that the irreducible matrix elements of $R^{[n]}$ are identical with the
original ones since $R^{[n]}$ is invariant under all symmetry transformations,
 i.e.\ it is a rank 0 object with respect to all symmetries.
Eq.~(\ref{eq:green_function_explicite}) explicitely shows 
that unless $A$ and $B$ have the same quantum numbers ({\underline a=\underline b}), 
$G^{R}_{A,B^{\dagger}}(z)=0$. 

\section{Numerical results}\label{sec:numres}

In this section we show the advantages of
using DM-NRG as opposed to NRG, and illustrate the benefits 
of using non-Abelian symmetries by applying DM-NRG to the 
two-channel Kondo (2CK) model. This model is exciting in itself 
as it possesses a non-Fermi liquid type of fixed point and it provides  
the simplest descriptions of the double dot 
system used recently to realize the 2CK state.\cite{Potok_07} 
This 2CK state is very fragile and in a  magnetic field the 
difference between the NRG and DM-NRG results are substantial. 

In the Wilson approach, the 2CKM is described by the following Hamiltonian, 
\begin{multline}
H^{}_{2CK} = \frac{1}{2}\;{\vec S}\sum_{\alpha \in\{1,2\}}{\cal J}^{}_{\alpha} 
\sum_{\sigma, \sigma^\prime \in \{\uparrow, \downarrow \}}
  f_{0, \alpha, \sigma}^{\dagger}{\vec \sigma^{}_{\sigma \sigma^\prime}}f^{}_{0,\alpha, \sigma^\prime} + \\
+\sum_{n=0}^{\infty}\;\sum_{\alpha\in\{1,2\}}\;\sum_{\sigma\in \{\uparrow, \downarrow\}}
t^{}_n \left ( f_{n,\alpha, \sigma}^{\dagger} f^{}_{n,\alpha, \sigma} + h.c.  \right)\; ,
\label{eq:2_chan_kondo_nrg}
\end{multline}
where we have introduced $\alpha\in\{1,2\}$ for labeling the two types of
electrons. This additional channel label is the only 
difference compared to the one-channel Kondo Hamiltonian (cf.\ Eq.\
(\ref{eq:1_chan_Kondo})). 

In this model, the number of carriers is conserved in both channels   
corresponding to a
$\textrm{U}_{C1}(1)\times\textrm{U}_{C2}(1)$ symmetry. 
However, due to the presence  of electron-hole symmetry, these 
charge symmetries are augmented to $\textrm{SU}(2)$ symmetries, 
and the  Hamiltonian above is also invariant under 
$\textrm{SU}_S(2)\times\textrm{SU}_{C1}(2)\times\textrm{SU}_{C2}(2)$
transformations. 
On the other hand, one can solve this Hamiltonian 
using exclusively $\textrm{U}(1)$ symmetries,
 $\textrm{U}_S(1)\times\textrm{U}_{C1}(1)\times\textrm{U}_{C2}(1)$. 
This model is thus ideal  for testing our flexible methods.

If a local magnetic field is coupled to the
impurity spin through a  term $g\mu_BBS^z$, then from among the total spin
generators (Eq.\ (\ref{eq:generators_spin})) 
 solely $S^z_T$ will commute with the
Hamiltonian. That is, the spin SU(2) symmetry of the system reduces 
to U(1).   Therefore, in a magnetic field  
we can either use the symmetry 
$\textrm{U}_S(1)\times\textrm{SU}_{C1}(2)\times\textrm{SU}_{C2}(2)$ for our
calculations, or restrict ourself to  $\textrm{U}(1)$ symmetries only:
$\textrm{U}_S(1)\times\textrm{U}_{C1}(2)\times\textrm{U}_{C2}(2)$.

As a test, we computed the  retarded Green's function, 
$G^{R}_{f_{0,\alpha,\uparrow}^{}, f_{0,\alpha,\uparrow}^{\dagger}}(\omega)$ 
 and the corresponding spectral function 
$\rho_{f_{0,\alpha,\uparrow}}(\omega)$  both in the
presence  and in the absence of magnetic field. 
All numerical results presented were obtained at zero temperature, and  
the dimensionless
couplings were ${\cal J}_\alpha=0.2$ for both channels and all runs. The
discretization parameter $\Lambda=2$ was used in all cases, and for each
symmetry combinations we have retained a maximum number of 1350 multiplets 
in each iteration.  

\begin{figure}[t]
  \includegraphics[width=0.9\columnwidth,clip]{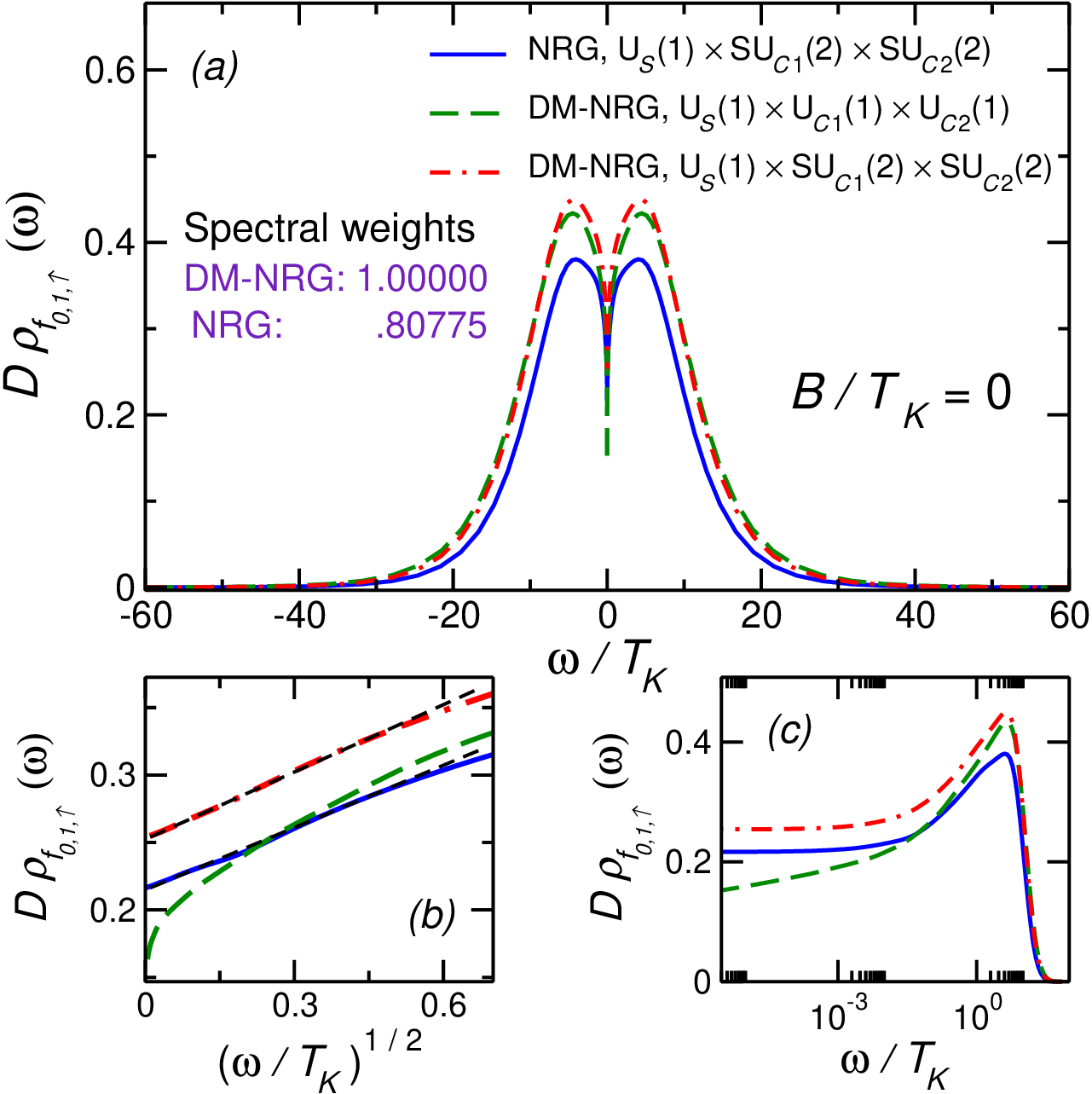}
  \caption{(color online) Dimensionless spectral function of $f_{0,1,\uparrow}$
    normalized by $D$, the bandwidth cut-off, as a function of $\omega/T_K$ in the
    absence of magnetic field obtained with DM-NRG and with NRG
    using the symmetries: $\textrm{U}_S(1)\times\textrm{SU}_{C1}(2)\times
    \textrm{SU}_{C2}(2)$ and $\textrm{U}_S(1)\times\textrm{U}_{C1}(1)\times
    \textrm{U}_{C2}(1)$.  $(a)$ Comparison between the spectral weights of the DM-NRG and NRG 
    results: DM-NRG fulfills the sum rule entirely even when the used symmetry
    group and therefore the number of kept states is largely reduced. NRG violates
    the sum rule to over $15\%$ if the number of kept states is
    $\approx7\times 10^4$ in  each iteration. $(b)$ The same spectral functions
    as a function of $\sqrt{\omega/T_K}$. If a sufficient number of
    states is kept, i.e. when using larger symmetry groups, 
    the expected $\sqrt\omega$ behavior around the 2CK fixed point is nicely
    recovered. $(c)$ The same spectral functions on a logarithmic scale.
  }\label{fig:b0}
\end{figure}

In Fig.\ \ref{fig:b0} we show data 
for the local fermion's spectral function in the absence of magnetic field 
obtained through the NRG and the DM-NRG approaches using the two symmetry
groups mentioned above.
The Kondo scale $T_K$ in Fig.\ \ref{fig:b0} is the scale at which  the 
2CK state forms, and it is defined as the frequency where the
$T$-matrix of the 2CK model drops to half of its value assumed at $\omega=0$
at the 2CK fixed point.\cite{2ck_cond} 

The first important 
test is the fulfillment of the spectral 
sum rules. These are always satisfied 
in the DM-NRG calculations independently of the symmetry group used, 
whereas the NRG data violate the sum rule to over 15$\%$ if the number
of kept multiplets is 1350 corresponding to $\approx7\times10^4$ states. 
Fig.\ \ref{fig:b0}.$(b)$ shows that the
expected $\sqrt\omega$ behavior around the 2CK fixed point is nicely 
recovered by both methods,  but  a sufficiently large number 
of multiplets must be kept in the DM-NRG approach, meaning that 
in this case  the larger symmetry group must be used. 
Fig.\ \ref{fig:b0}.$(c)$ demonstrates that, in spite of 
fulfilling the spectral sum rules still the DM-NRG data do not show
the expected asymptotics for low frequencies if the number of multiplets kept
is not sufficient. It is quite remarkable that only the DM-NRG procedure 
using non-Abelian symmetries was able to get close 
 to the exact value of the spectral function at $\omega=0$, 
$\rho_{f_{0,\alpha,\uparrow}}(\omega=0) = 0.25$.

The presence of magnetic field generates a new scale,\cite{Andrei_84} 
\bea
T_h\equiv C_h\; \frac{B^2}{T_K}\;,
\eea 
where we have fixed the somewhat arbitrary constant to 
$C_h\approx 60$.\cite{2ck_spf} This scale is usually
referred to as the renormalized magnetic field acting on the impurity, and
below this scale the non-Fermi liquid physics is destroyed. 

In the presence of magnetic field the sum rule is violated 
by the NRG approach 
to a different extent in the positive and negative frequency ranges, 
which leads
 to jumps at $\omega=0$ in the spectral functions (see Fig.\
\ref{fig:bneq0}), while this problem is absent in the DM-NRG
approach. 

\begin{figure}[t]
  \includegraphics[width=0.9\columnwidth,clip]{figs/bneq0}
  \caption{(color online) Dimensionless spectral function of $f_{0,1,\uparrow}$
    normalized by $D$, the bandwidth cut-off, as a function of $\omega/T_K$ in the
    presence of magnetic field obtained with DM-NRG and with NRG
    using the symmetries: $\textrm{U}_S(1)\times\textrm{SU}_{C1}(2)\times
    \textrm{SU}_{C2}(2)$ and $\textrm{U}_S(1)\times\textrm{U}_{C1}(1)\times
    \textrm{U}_{C2}(1)$. $(b)$ On a smaller scale at $\omega=0$ we show the smoothness 
    of the DM-NRG data using both groups and the jump in the NRG
    results using the larger group.
  }\label{fig:bneq0}
\end{figure}

\begin{figure}[t]
  \includegraphics[width=0.9\columnwidth,clip]{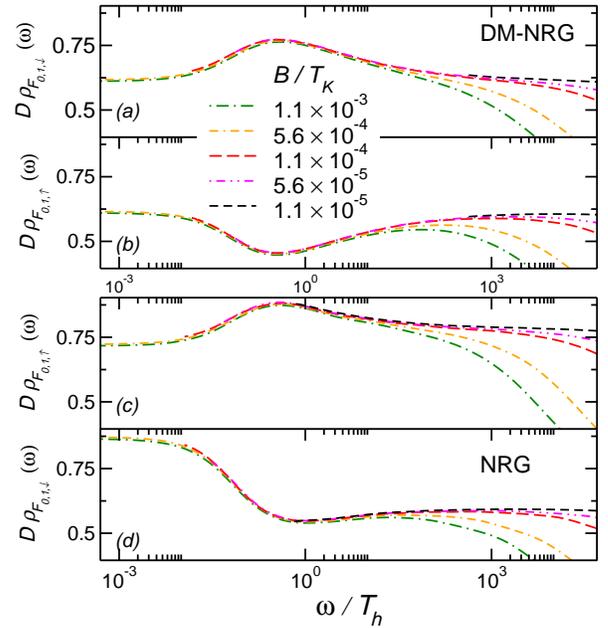}
  \caption{(color online) Dimensionless spectral function of the $\downarrow$- $(a),(c)$ and the
    $\uparrow$-spin $(b),(d)$ components of the local composite fermion
    operator normalized by $D$, the bandwidth cut-off,  
    for sufficiently small values of $B$ as a function of $\omega/T_h$ scaled 
    on top of each other using NRG and DM-NRG together with the group
    $\textrm{U}_S(1)\times\textrm{SU}_{C1}(2)\times\textrm{SU}_{C2}(2)$. 
  }\label{fig:cf_scaled_dmnrg_vs_nrg}
\end{figure}

\begin{figure}[b]
  \includegraphics[width=0.9\columnwidth,clip]{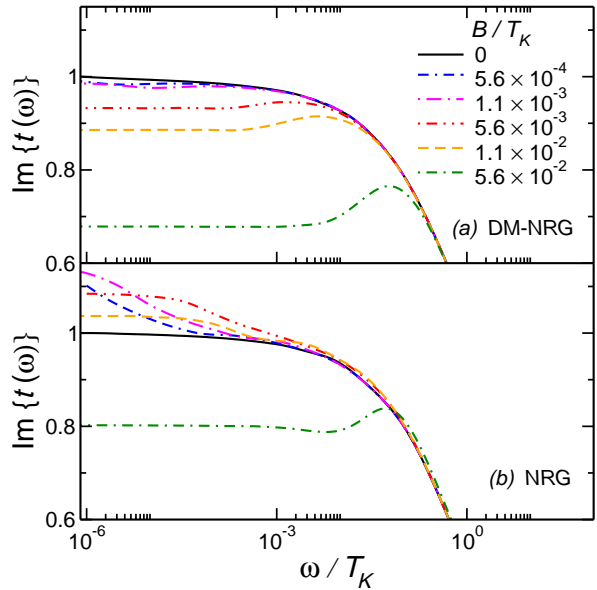}
  \caption{(color online)
    Imaginary part of the on-shell $T$-matrix using 
    DM-NRG and NRG together with the group $\textrm{U}_S(1)\times\textrm{SU}_{C1}(2)\times
    \textrm{SU}_{C2}(2)$ for various magnetic field values as a function of 
    $\omega/T_K$. 
  }\label{fig:im_t_nrg_vs_dmnrg}
\end{figure}

Spectral functions display universal scaling in the vicinity of 
$T_h$.\cite{2ck_spf}
In Fig.\ \ref{fig:cf_scaled_dmnrg_vs_nrg} we show
how  the spectral
functions of the composite fermion operator, 
$F^\dagger_{1}\equiv f^\dagger_{0,1} \vec S \vec \sigma$ can be scaled 
on top of each other using the scale $T_h$. 
Although, this collapse can be obtained in both approaches,  
there is an $\approx20\%$ jump at $\omega=0$ in the NRG results while 
the DM-NRG results are continuous there. 
It is possible to eliminate the jump in the NRG results by determining the
phase shifts from the energy spectrum with high precision, but even after 
 these
corrections, the results continue to violate the sum rule and numerical
errors for low frequencies remain of the same size as before.
Moreover, the 'universal' curve
obtained by conventional NRG is clearly incorrect and 
different from the DM-NRG  result. 
Also, if we try to   compute the imaginary part of the on-shell $T$-matrix 
of the 2CK model where the local composite fermion spectral 
functions for both $\uparrow$-
and $\downarrow$-spin components have to be summed up, we end up
with large numerical errors in the NRG results (see Fig.~\ref{fig:im_t_nrg_vs_dmnrg}), while DM-NRG provides satisfactory 
results even in this case.
It is thus clear from these examples that the DM-NRG method
together with the use of lots of symmetries produces much more 
reliable results than NRG with non-Abelian symmetries 
or DM-NRG with only Abelian symmetries, and its use is needed to do
computations for more delicate quantum impurity models.

\section{Conclusions}\label{sec:conclusion}

To summarize, in this paper, we have shown how the recently developed spectral sum-conserving DM-NRG
methods can be used in the presence of any number and type of non-Abelian symmetries.
The most important result of this paper is a very general and simple
recursion relation for the truncated reduced  density matrix, which
enables one to compute the spectral properties of any local correlation function
in the presence of non-Abelian symmetries.
The expressions derived hold for almost any symmetry, including non-Abelian finite
groups, point groups, SU(N) groups, and, of course,  Abelian groups such as
U(1) or the parity.
Using  these symmetries, reduces considerably the time needed for
the computations  and enhances significantly the accuracy of the
numerical results for dynamical quantities. The general formulation  
presented in this paper also allowed us to construct
a flexible DM-NRG code where symmetries are handled dynamically,
and which is able to learn and  handle essentially any type of symmetry.\footnote{This
code has open access and, together with a detailed description, it will be accessible
soon  at the site  http://www.phy.bme.hu/\textasciitilde dmnrg.}

We demonstrated the advantages of the generalized method by applying it  
to the two-channel Kondo model in a magnetic field.
The presence of magnetic field makes this model very challenging from the
point of view of NRG calculations. We have carried out calculations for the
local fermion's and for the local composite fermion's spectral function
at zero temperature and in a finite magnetic field. In conventional
NRG, there is always a jump  at $\omega\to0$ between the positive and negative frequency
parts of the spectral function. In addition to being spectral sum-conserving,
the use of DM-NRG almost completely eliminated this jump  and made it possible to  compute
the universal cross-over curves in a magnetic field.  Moreover,  the DM-NRG approach used with
larger symmetry groups has provided substantially better results for
the spectral functions: In the limit $\omega\to0$ we needed to use non-Abelian
symmetries in the DM-NRG approach to recover  the expected power law behavior known
from conformal field theory. Also, the universal crossover curve and the shape of the
the peak at the renormalized magnetic field $T_h\propto B^2/T_K$
was much better resolved by DM-NRG than by NRG, and thus DM-NRG with non-Abelian symmetries
lead  to much  more reliable results for the $T$-matrix (see Fig.\ \ref{fig:cf_scaled_dmnrg_vs_nrg})
than NRG with non-Abelian  or DM-NRG with Abelian symmetries.
We emphasize that, to obtain the $\omega=0$ value of the spectral function correctly
as well as its proper scaling behavior, we needed to use non-Abelian symmetries.

The method presented here thus opens up the possibility to carry out very accurate
spectral sum-conserving DM-NRG calculations for multichannel systems,
such as multi-dot devices, and to  perform reliable DM-NRG-DMFT calculations for multichannel lattice
models. Combined with the matrix product state approach, it might also provide a way to
use methods applied in DMRG to improve the high frequency resolution of NRG.

\acknowledgments
We are deeply indebted to Ireneusz Weymann, who participated in the initial
stage of the construction of the flexible NRG code, and to L\'aszl\'o Borda for
his help and many comments. We would like to thank Frithjof  Anders, Jan von
Delft and Andreas Weichselbaum for useful discussions.
This research was supported partly by the Hungarian Research Fund (OTKA)
under Grant Nos. NF 61726 and K 68340 and by the J\'anos Bolyai Research
Fund.   A.T.\ is grateful to the Institute of Mathematics of the
BUTE for providing  access to their computer cluster, supported by OTKA under
grant No. 63066. CPM was partially supported by the Romanian Grant No. CNCSIS 780/2007.

\section{Proof of the diagonal form of the reduced density matrix
for $\textrm{SU(2)}$ symmetries}\label{app:SU2}

In this appendix we prove, in a different way from how it was demonstrated in the main
part of the article, the general 
theorem for the diagonal form of the reduced density matrix presented in Section \ref{sec:construction}
in the case when only $\rm SU(2)$ \emph{local symmetries} are involved. 

The proof for the diagonal form of the reduced density matrix goes by induction for the iteration
steps. At the last iteration, by construction the reduced density matrix $R^{[N]}$ is
scalar under the local symmetry group (cf.\ Eq.\ (\ref{eq:recursive})). 
Due to this invariance by very general considerations  
$R^{[N]}$ can be written as a sum over tensor products of irreducible tensor
operator components, and is  of the form
\bea\label{eq:scalar_op_decomp}
R^{[N]}=\ds{\sum_\alpha}\left(T^{[loc]}_\alpha\right)_{{\underline t},{\underline
    t}^z}^\dagger\otimes 
\left(T^{[N-1]}_\alpha\right)_{{\underline t},
{\underline t}^z}\;,
\eea
where $\left(T^{[loc]}_\alpha\right)_{{\underline t},{\underline
    t}^z}$ and $\left(T^{[N-1]}_\alpha\right)_{{\underline t},
{\underline t}^z}$  
are irreducible tensor operator components of the same rank ${\underline t}$ 
acting on the local vector space at site $N$ and on the
rest of the Wilson chain, respectively and $\alpha$ labels
all possible tensor operators. 
This special form is a consequence of
the invariance of $R^{[N]}$ under the local symmetry transformations.\cite{thesis}

To obtain $R^{[N-1]}_{KK}$ we have to
trace over the local basis states, $\left\{\left|\mu, {\underline
  q}_\mu,{\underline q}^z_\mu\right>\right\}$, 
that is we have to compute the following matrix elements
\begin{multline}\label{eq:central}
\left<u \right . \parallel R^{[N-1]}_{KK}\parallel\left . v\right>
=\ds{\sum_{\alpha,{\underline
      q}_{\mu},{\mu}}}\textrm{sgn}\left(T^{[N-1]}_\alpha,{\underline
      q}_{\mu}\right)\\\times
\left<u\right . \parallel
  T_\alpha^{[N-1]}\parallel\left . v\right>
\left<\mu \right .\parallel
  T^{[loc]}_\alpha\parallel\left . \mu\right>\\
\times\left<\underline{Q}_u\;
\underline{Q}^z_u\left|\underline{t}\;\underline{t}^{z}\;\underline{Q}^{}_v\;
\underline{Q}^z_v\right>\right.\\
\times\ds{\sum^{\underline q_{\mu}}_{
 \underline q^{z}_{\mu}=-\underline q_{\mu}}}\left< \underline q_{\mu}
\underline q^z_{\mu}\left|\underline t\, \underline t^{z}\,
\underline q_{\mu}\, \underline q^z_{\mu}\right>\right.\;.
\end{multline}

In Eq.\ (\ref{eq:central}) we have applied the Wigner--Eckart theorem.
The  sign factor $\textrm{sgn}(.,.)$ depends on the number of fermionic operators 
used in the construction of $T^{[N-1]}_\alpha$ and the local states.
In Eq.\ (\ref{eq:central}) in the last sum only the terms with 
$\underline t_z =0$ give  non-vanishing contributions and the sum can be
reduced to   
\begin{equation}
\ds{\sum_{q^z=-q}^q}
\left<q\;q^z\left|t\;0\;q\;q^z\right>\right.=\left(2q+1\right)\delta_{t,0}\;.\label{eq:last}
\end{equation}

Eq.\ (\ref{eq:last}) implies that the only non-vanishing contributions 
are those corresponding to $\underline t =\underline t^z= 0$, i.e. 
to scalar $T^{[loc]}_\alpha$ and  $T_\alpha^{[N-1]}$. 
Therefore the reduced density matrix 
$R^{[N-1]}$ is diagonal in the representation indices. 
The induction towards smaller iteration steps goes recursively the same way.

\end{document}